\documentclass[oupdraft]{bio}
\usepackage{amsthm,amsmath,amssymb,graphicx,setspace,verbatim}
\usepackage{subcaption}
\usepackage{wrapfig}
\usepackage{multirow}
\usepackage{rotating}
\usepackage{color}
\usepackage{adjustbox}
\usepackage{changepage}

\newcommand{\scriptsize}{\footnotesize}

\renewcommand{\arraystretch}{.95}

\newcommand{\bx}{\mbox{\boldmath $x$}}

\newcommand{\by}{\mbox{\boldmath $y$}}

\newcommand{\bI}{\mbox{\boldmath $I$}}

\newcommand{\bP}{\mbox{\boldmath $P$}}

\newcommand{\balpha}{\mbox{\boldmath $\alpha$}}
\newcommand{\bbeta}{\mbox{\boldmath $\beta$}}

\newcommand{\bgamma}{\mbox{\boldmath $\gamma$}}
\newcommand{\bdelta}{\mbox{\boldmath $\delta$}}

\newcommand{\bSigma}{\mbox{\boldmath $\Sigma$}}

\newcommand{\bPsi}{\mbox{\boldmath $\Psi$}}
\newcommand{\bPhi}{\mbox{\boldmath $\Phi$}}

\newcommand{\bnu}{\mbox{\boldmath $\nu$}}

\newcommand{\Cov}{\mbox{Cov}}

\newcommand{\bdm}{\begin{displaymath}}
\newcommand{\edm}{\end{displaymath}}
\newcommand{\beq}{\begin{equation}}
\newcommand{\eeq}{\end{equation}}

\renewcommand{\th}{^{\mbox{\scriptsize th}}}

\long\def\symbolfootnote[#1]#2{\begingroup%
\def\thefootnote{\fnsymbol{footnote}}\footnote[#1]{#2}\endgroup}

\newlength{\offsetpage}
\newenvironment{widepage}{\begin{adjustwidth}{-\offsetpage}{-\offsetpage}%
    \addtolength{\textwidth}{2\offsetpage}}%
{\end{adjustwidth}}

\begin{document}

\title{\vspace{-.6in}Prediction of Individual Outcomes for Asthma Sufferers\vspace{-.1in}}

% List of authors, with corresponding author marked by asterisk
\author{CURTIS B.~STORLIE${^*}^\dag$,  MEGAN E.~BRANDA$^\dag$, MICHAEL R.~GIONFRIDDO$^\ddag$, \\[-3pt]
  NILAY D.~SHAH$^\dag$, MATTHEW A.~RANK$^\dag$ \\[8pt]
% Author addresses
\textit{
  $\dag$Mayo Clinic, Rochester, MN\\[-5pt]
  $^\ddag$Geisinger Health System, Danville, PA
  %$^\S$Geisinger Health System\\  %Center for Pharmacy Innovation and Outcomes,
}
%\\[2pt]
% E-mail address for correspondence
%{storlie.curt@mayo.edu}
}

% Running headers of paper:
\markboth%
% First field is the short list of authors
{C.~B.~Storlie et.~al.}
% Second field is the short title of the paper
{Prediction of Individual Outcomes for Asthma Sufferers}

\maketitle

%\footnotetext{blah}

\begin{abstract}
{We consider the problem of individual-specific medication level recommendation (initiation, removal, increase, or decrease) for asthma sufferers.  Asthma is one of the most common chronic diseases in both adults and children, affecting 8\% of the US population and costing \$37-63 billion/year in the US.
  %Trend analyses of population asthma health outcomes over the past decade suggest that the costs of asthma care are increasing (primarily driven by medication costs) and that asthma outcomes have not significantly improved.
  %This implies that there may be an opportunity to provide a more optimal alignment of asthma medication to a given patient.
  Asthma is a complex disease, whose symptoms may wax and wane, making it difficult for clinicians to predict outcomes and prognosis. Improved ability to predict prognosis can inform decision making and may promote conversations between clinician and provider around optimizing medication therapy.  Data from the US Medical Expenditure Panel Survey (MEPS) years 2000-2010 were used to fit a longitudinal model for a multivariate response of adverse events (Emergency Department or In-patient visits, excessive rescue inhaler use, and oral steroid use).  To reduce bias in the estimation of medication effects, medication level was treated as a latent process which was restricted to be consistent with prescription refill data.  This approach is demonstrated to be effective in the MEPS cohort via predictions on a validation hold out set and a synthetic data simulation study.  This framework can be easily generalized to medication decisions for other conditions as well.}{Asthma; Hierarchical Bayesian modeling; Latent process; Log-Gaussian Cox process; Measurement Error Model; Medication step-down.}
\end{abstract}

%\begin{keywords}
%Asthma; Hierarchical Bayesian modeling; Latent process; Log-Gaussian Cox process; Measurement Error Model; Medication step-down.
%\end{keywords}

\section{Introduction}
Asthma is a chronic disease, affecting 8\% of the population and costing \$37-63 billion/year in the US \citep*{Kamble2009,Jang2013}.  Asthma can lead to death and also includes significant morbidity including hospitalization, emergency department visits, outpatient visits, missed school and work days, decreased productivity, and decreased physical activity.
%is one of the most common chronic diseases in both adults and children and the disease burden associated with asthma can include death, hospitalizations, emergency department visits, outpatient visits, missed school and work days, decreased productivity, and decreased physical activity.
%There is no cure for asthma, and no sure-fire method to prevent asthma from developing.
Current strategies for managing asthma include avoidance of triggers, anticipatory guidance for exacerbation treatment, and the scheduled use of various asthma medications as “preventive” medications.
Medications account for the highest proportion of healthcare spending for asthma.  In addition, spending on medications has been growing at a rapid rate over the last decade \citep*{Rank2012a}.
%Medication represents the single largest direct cost category of asthma management, and is also the category that is increasing the most over time \citep*{Rank2012a}.
While medication spending continues to increase, this  has not resulted in improved asthma outcomes \citep*{Moorman2012,Rank2012b}.
%Trend analyses of population asthma health outcomes also suggest that asthma outcomes have not significantly improved \citep*{Moorman2012,Rank2012b}.
To enhance the management of asthma, we propose to develop a risk prediction model to assist clinicians and patients when deciding how best to manage asthma medication for a given patient.
%In an effort to improve outcomes and lower cost, we seek to develop a principled approach to assist clinicians and patients when deciding how best to manage asthma medication for a given patient.

The current strategy of asthma medication management includes an assessment of current symptom control and future risk for asthma exacerbations which is then matched for treatment intensity.
%The treatment intensity is generally depicted in asthma guidelines as “step-level” care based on the type and dose of asthma controller medications used.
Assessing current symptoms can be accomplished by using standardized patient-reported tools; however, estimating future risk is more challenging.  Prior efforts to predict future exacerbation events focused on specific risk factors available in cross-section (e.g. lung function, blood eosinophil levels, prior exacerbations, or a combination of these and other factors) \citep*{Bateman2015,Blakey2012}.  However, each individual with asthma may have a differential response to medications over time.  Using this individual response data to predict outcomes at differing medication levels has not been previously attempted.  Our hypothesis is that using response to medication data at the individual level will provide a meaningful future risk estimate.  Individualized estimation of future risk will provide a key step in personalizing asthma medication regimens.
%for individuals with this common and chronic disease.

A total of $4,235$ patients with persistent asthma were identified from the US Medical Expenditure Panel Survey (MEPS) years 2000-2010. Each patient had data for $\sim$2 years, and measurement was divided into 5 rounds (periods of 4 to 5 months each). Medication level, taking on ordinal levels of 0 $-$ 5, was defined by type and dose of chronic asthma medication based on current guidelines as described in \citet*{Rank2016}.  This definition of medication level only includes controller medication and does not include rescue inhaler or oral steroid usage which are both considered outcomes here.  The primary outcomes of interest are (i) emergency department (ED) and/or in-patient (IP) visits, (ii) Rescue inhaler refills, and (iii) Oral steroid use.  The goal of this work is to provide an individualized risk prediction of an {\em adverse event} in the next six months for a given patient on a given medication level.  An adverse event is defined here to be any of the following: (i) any ED/IP visit, (ii) any oral steroid use, (iii) four or more rescue inhaler refills.

Unfortunately in the MEPS data, medication level is observed only as an aggregate (i.e., the average level) for an entire round.  This is problematic because in reality medications are often introduced or increased at an ED/IP visit (which often occur in the middle of a round).  If this were to be ignored and medication level were assumed constant throughout a round, then the MEPS data would actually imply that higher medication levels for a round are associated with a higher chance of an ED/IP visit.  Thus, some care is needed when modeling the effect of medication level.  We propose a {\em latent} random process for the medication level for an individual so that it is allowed to change within a round, with a positive probability of changing at the time of an IP, ED, office-based (OB), or out-patient (OP) visit.  This process is restricted so that the average medication level for an entire round is consistent with the observed, average medication level for that round in the MEPS data.  Latent variable approaches are common in the literature and have been used in similar situations \citep*[e.g.,][]{Muthen83,Dunson2000,Storlie12a}.  Conditional on the latent medication process, the model for the multivariate response over time is a dependent log-Gaussian Cox Process \citep*{moller1998log,brix2001spatiotemporal} with a mean function that depends on a patient's covariates and a random intercept.  Thus, this model can be considered as a measurement error model \citep*{carroll2006} on a time-varying covariate \citep*{wulfsohn1997,tsiatis2001}.  The unique features here are that the time varying covariate cannot be assumed to be a smooth process over time and the variable is never measured/known exactly at any particular times. In this case, only the average level is known over several partitions of time and it is quite possible for the process to jump from one level to another.  Estimation is performed in a Bayesian framework via Markov Chain Monte Carlo (MCMC) in a similar manner to that for handling multivariate missing data \citep*{Storlie17a,Storlie17b}. 

The remainder of the paper is laid out as follows.  Section~\ref{sec:model} describes the statistical approach used to model patient outcomes, while computational details are discussed in Section~\ref{sec:computation}.  Analysis of the MEPS data and a summary of predictive accuracy is provided in Section~\ref{sec:MEPS}.  Section~\ref{sec:sims} summarizes the results of applying the model to a simulated test case designed to be similar to the the MEPS data and Section~\ref{sec:conclusions} concludes the paper.

\section{Statistical Model}
\label{sec:model}
The outcomes of rescue inhaler usage and oral steroid usage are aggregated at the round level.
%as there is no way to know when in the round they were filled in the MEPS data.
However, ED/IP visits are captured at the day granularity, so the model below is defined on a finer granularity of {\em time period} within each round.  In principle the time period could be as fine as one day, however, one week granularity was used in all results below for computational expediency.  The response for the $i\th$ individual for period $n$ (i.e., week $n$) is $\by_{i,n} = [y_{i,n,1}, y_{i,n,2}, y_{i,n,3}]'$, with
\vspace{-.1in}\begin{eqnarray}
y_{i,n,1} & = & \mbox{\# of ED/IP visits for $i\th$ individual during $n\th$ time period.} \nonumber\\
y_{i,n,2} & = & \mbox{\# of rescue inhaler refills for $i\th$ individual during $n\th$ time period.} \label{eq:responses}\\
y_{i,n,3} & = & \mbox{\# of oral steroid refills for $i\th$ individual during $n\th$ time period.} \nonumber\\[-.45in] \nonumber
\end{eqnarray}
As discussed above, the $y_{i,n,1}$ are actually observed, but only the sum of $y_{i,n,2}$ and $y_{i,n,3}$ for the $m\th$ {\bf round} are observed, i.e.,
\vspace{-.1in}\begin{eqnarray}
y^*_{i,m,2} &=& \sum_{n \in R_{i,m}} y_{i,n,2}, \nonumber\\
y^*_{i,m,3} &=& \sum_{n \in R_{i,m}} y_{i,n,3}, \nonumber\\[-.35in] \nonumber
\end{eqnarray}
where $R_{i,m}$ is the set of time periods that make up the $m\th$ round.

The distribution of these responses was assumed to be a function of patient dependent covariates $\bx_{i} = [x_{i,1},\dots,x_{i,p}]'$ (including physical characteristics, co-morbidities, etc.) and their asthma controller medication level $z_i(t)$ at time $t$.  The controller medication level takes on ordinal levels of 0 $-$ 5 as defined in \citet*{Rank2016}.  In the MEPS data, $z_i(t)$ is observed only as an aggregate (i.e., the average level) for an entire round.  Hence, the $z_i(t)$ were assumed here to be (latent) random processes so that they can be allowed to change within a round.  It is essential to allow for this in order to remove the bias introduced by the fact that medications are often introduced and/or increased at an ED/IP visit.  If this were to be ignored in a model, then it would incorrectly imply that higher medication levels increase the chance of an ED/IP visit.  Thus, $z_i$ was allowed to be a Markov process with the restriction that the average medication level for an entire round must be consistent with the observed (aggregated) medication level for that round in the MEPS data.  There is a baseline rate at which increases or decreases to medication level may occur at any given time, but there is also a probability $\varpi$ of changing immediately after a IP/OP/OB/ED visit (i.e., the probability of a change in a given week is allowed higher if there was contact with a provider during that week).  The specifics of the model used for controller medication level are described later in this section.  Identifiability concerns are addressed in Section~\ref{sec:computation}.

The distribution of $\by_{i,n}$ also depends on a time varying random effect for the $i\th$ individual to account for the overall severity of their condition compared to others in the population and the variation in the severity of an individual's symptoms over time.  %The distribution of $\by_{i,n}$ should also clearly be dependent on the length of the $n\th$ time period.
A model to account for all of this is the following hierarchical Poisson process,
\beq
y_{i,n,l} \sim \mbox{Poisson}(\lambda_{i,n,l}),
\label{eq:model_1}
\eeq
where
\vspace{-.0in}\bdm
\lambda_{i,n,l} = \int_{t_{n-1}}^{t_n} \eta_{i,l}(s) ds,
\vspace{-.0in}\edm
where $t_n$ is the end of the $n\th$ time period and
\vspace{-.05in}\beq
\log \left[ \eta_{i,l}(s) \right]= \alpha_{i,l}  + \sum_{j=1}^J \beta_{j,l} x_{i,j} + \gamma_l z_i(s) + \delta_{i,l}(s),
\label{eq:model_2}
\vspace{-.05in}\eeq
where $l=1,2,3$ is indexing the response type (i.e., ED/IP visits, rescue inhaler usage, and oral steroid usage).
\vspace{-.1in}\begin{itemize}
  \item $\alpha_{i,l}$ is the random effect on the $l\th$ response type, to account for the overall level of severity for the $i\th$ individual and for the individual's response to medication.
  \item $\beta_{j,l}$ are fixed effects, constant across the population.
  \item $\gamma_l$ is the fixed effect for controller medication level on the $l\th$ response type.
  \item $\delta_{i,l}(s)$ is a time varying random effect to account for the natural fluctuation/trends in an individual's condition. 
\vspace{-.1in}\end{itemize}

This model implies that the round aggregated responses follow,
\vspace{-.0in}\begin{eqnarray}
y^*_{i,m,2} \sim \mbox{Poisson}\left(\sum_{n \in R_{i,m}} \! \lambda_{i,n,2}\right), \nonumber \\
y^*_{i,m,3} \sim \mbox{Poisson}\left(\sum_{n \in R_{i,m}} \! \lambda_{i,n,3}\right). \nonumber \\[-.35in] \nonumber
\end{eqnarray}

In order to complete the model specification a parent distribution must be defined for the random effects $\alpha_{i,l}$ and $\delta_{i,l}$.  A multivariate normal distribution was assumed for the $\alpha_{i,l}$,
\vspace{-.1in}\beq
\balpha_i = \left[\alpha_{i,1},\dots,\alpha_{i,L} \right]'  \sim  N \left( \bnu, \bSigma \right).
\label{eq:model_3}
\vspace{-.1in}\eeq

The time varying effect $\delta_{i,l}$ was assumed to be a Gaussian process with mean 0 and covariance function $K_l(s,t)$; specifically, it is a stationary Ornstein-Uhlenbeck process (equivalent to an order one, autoregressive time series) for each $l$, i.e.,
\vspace{-.1in}\bdm
K_l(s,t) = \tau_l \exp \{-\theta |s-t|\}.
\vspace{-.1in}\edm
A product correlation was also assumed between $\delta_{i,l}$  and $\delta_{i,l'}$ so that the variation in the rate of occurrence for the three response types are correlated, i.e.,
\vspace{-.1in}\bdm
\Cov \left( \left[\delta_{i,1}(s), \dots, \delta_{i,L}(s) \right]' \right) = \bPhi, 
\vspace{-.1in}\edm
or the general covariance function is then 
%$$
%\Cov \left( \delta_{i,l}(s), \delta_{i,l'}(t) \right) = \left[I_{\{l=l'\}} + \rho I_{\{l \neq l'\}}  \right] \sqrt{\tau_l \tau_{l'}} \exp \{-\theta |s-t|\}.
%$$
\vspace{-.1in}\beq
K\left((l,s),(l',t) \right) = \Cov \left( \delta_{i,l}(s), \delta_{i,l'}(t) \right) =  \bPhi_{l,l'} \exp \{-\theta |s-t|\}.
\label{eq:model_4}
\vspace{-.1in}\eeq

With the above model, for a given individual and their response history, predictions can be made for risk of future outcomes at various levels of medication.  This prediction is a (correlated) distribution for the three responses so medication decisions can be assessed using the probability that the patient would need fewer than 4 rescue inhalers, not have an ED/IP visit, and not need OCS in the next six months, for example.  This probability or some similar measure can be used to assess if this individual is a good candidate for a step down in medication.

Finally, a model for the controller medication level, $z_i(t)$, is specified as
\vspace{-.05in}\beq
z_i(t) = \sum_{k=0}^\infty [\mu_h + \tilde{z}_{i,h}(t)] I_{\{B_i(t)=k\}} I_{\{\mu_h+\tilde{z}_{i,h}(t) > 0\}},
\label{eq:model_5}
\vspace{-.05in}\eeq
where (i) $\tilde{z}_{i,h}$, $h=0,1,\dots$, are {\em iid} mean zero Ornstein-Uhlenbeck (O-U) processes with variance $\sigma$ and correlation parameter $\phi$, (ii)  $I_A = 1$ if $A$, and 0 otherwise (e.g., the last term in (\ref{eq:model_5}) ensures that $z_i(t)\geq0$), (iii) $\mu_h \sim N(M_\mu, S^2_\mu)$, and (iv) the $B_i(t)$ are independent counting process intended to allow medication level to make large (discontinuous) changes at certain points in time.  That is,
\vspace{-.15in}\bdm
B_i(t) = C_i(t) + D_i(t),
\vspace{-.15in}\edm
where the $C_i$ allow the possibility of abrupt changes in medication level at any time, while the $D_i$ allow for a positive (increased) probability of an abrupt change at the time of a provider contact.  Specifically, it was assumed that $C_i$ is a Poisson process with intensity $\rho$ and $D_i(t)$ is a counting process that is allowed to increment by one, only at the time of an IP/OP/OB/ED visit.  That is, $D_i(0)=0$ and $D_i(t^+) = \lim_{s\downarrow t}D_i(s) = D_i(t)$ for all $t$, unless $t$ is the time of an IP/OP/OB/ED visit, in which case
\vspace{-.1in}\bdm
\begin{array}{rcl}
\Pr \{ D_i(t^+) = D_i(t)+1 \} & = & \varpi, \\[.1in]
\Pr \{ D_i(t^+) = D_i(t) \} & = & 1- \varpi.
\end{array}
\vspace{-.0in}\edm

%\vspace{-.1in}\bdm
%\Pr \{ D_i(t^+) = h \} = \left\{
%\begin{array}{ll}
%  \varpi & \mbox{for $h= D_i(t) + 1$,} \\
%  1-\varpi & \mbox{for $h= D_i(t)$,} \\
%  0 & \mbox{otherwise.}
%\end{array}
%\right.
%\vspace{-.1in}\edm

%\vspace{-.1in}\bdm
%\Pr \{ D_i(t^+) = h \} = \left\{
%\begin{array}{ll}
%  1-\varpi & \mbox{for $h= D_i(t)$ if $t$ is the time of an IP/OP/OB/ED visit,} \\
%  \varpi & \mbox{for $h= D_i(t) + 1$ if $t$ is the time of an IP/OP/OB/ED visit,} \\
%  1 & \mbox{for $h= D_i(t)$ if $t$ is not a time of an IP/OP/OB/ED visit,} \\
%  0 & \mbox{otherwise.}
%\end{array}
%\right.
%\vspace{-.1in}\edm

As mentioned above, the aggregated medication level for a round provided by the MEPS data $z^*_{i,m}$ was assumed to be the average of $z_i(t)$ over the $m\th$ round.  In the above model, the latent variable $z_i$ is continuous even though the observed value for the $m\th$ round, $z^*_{i,m}$ is ordinal on $0,\dots,5$.  This implies that the underlying medication level is continuous, but it is observed as an integer as a result of interval censoring on the cutpoints 0.5, 1.5, 2.5, 3.5, 4.5.  For example, an average medication level over round $m$ may be 2.3 or 6.2, which would be ``observed'' as a medication levels of 2 and 5, respectively.  Thus, the relation from $z_i(t)$ to $z^*_{i,m}$ is
\bdm
z^*_{i,m} = \min\left\{ 5\;,\; \left\lfloor \int_{t \in R_{i,m}} z_i(t)dt + 0.5 \right\rfloor  \right\}.
\edm

\begin{figure}[t!]
\vspace{-.13in}
\centering
\caption{25 posterior realizations of the latent controller medication process, $z_i$ for a given patient.  When an increase in medication occurs, the model generally prefers to have it occur at ED/IP/OP/OB visits over other times.}
\label{fig:z_i}
\vspace{-.4in}
\includegraphics[width=.85\textwidth]{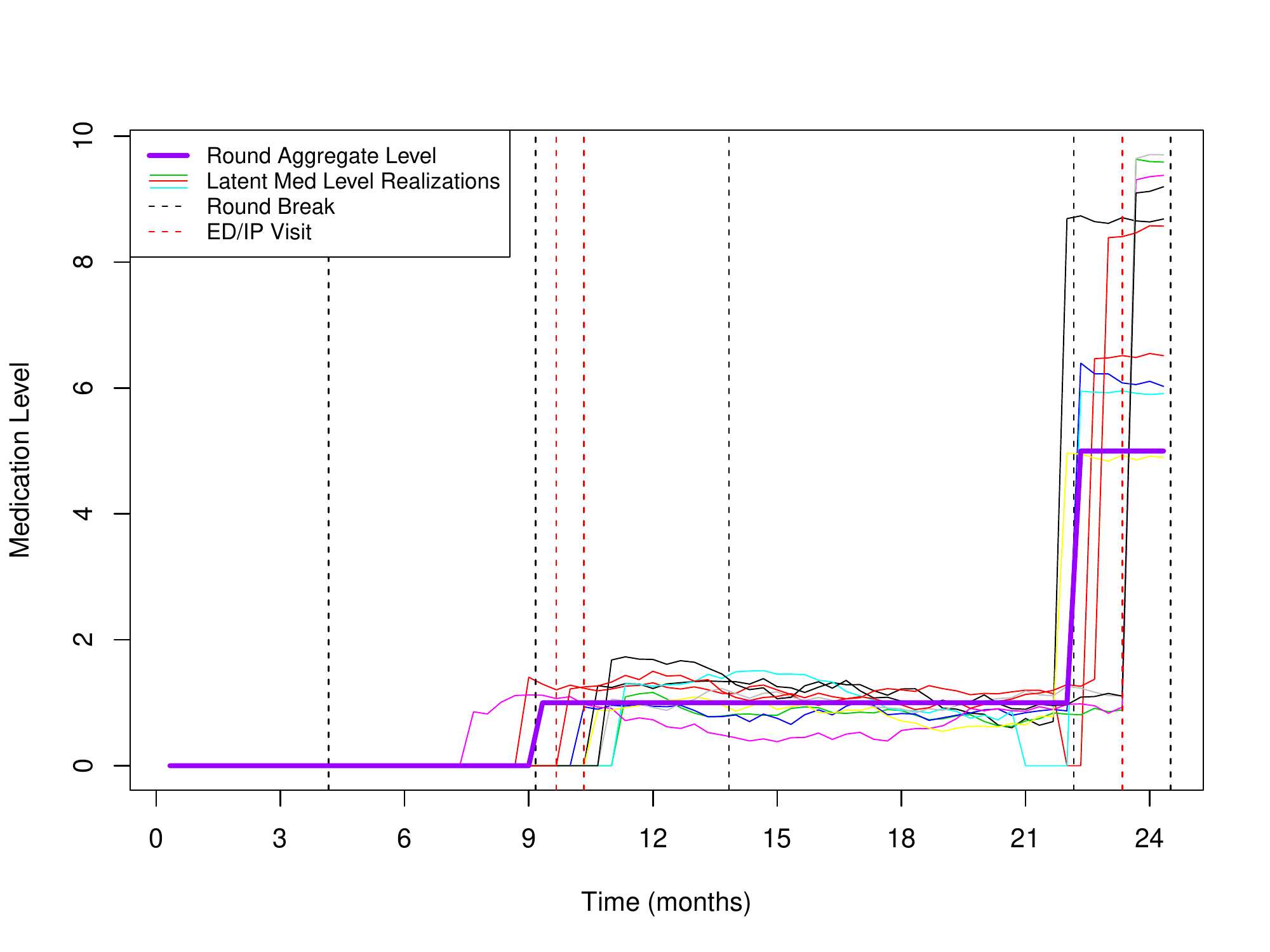}
\vspace{-.0in}
\end{figure}

Several realizations of $z_i$ resulting from posterior distribution for an example patient are provided in Figure~\ref{fig:z_i} along with the corresponding $z^*_{i,m}$ for reference.  The process described above for $z_i$ is Markov, which is appropriate for this application.  However, there may be situations where the medication level has a strong dependency on the entire history of previous medications.  Intuitively this should not have a large influence on the main goal is {\em interpolation} of medication level as opposed to prediction into the future.  However, there may be situations where this dependence to prior history is very strong and could be leveraged into a better model, perhaps via other covariance functions (e.g., powered exponential or Matern) and/or non-Markovian counting processes, etc.

\section{Prior Distributions and Computation}
\label{sec:computation}
A summary of the model structure and the assumed prior distributions are provided in Table~\ref{tab:priors}.  Relatively diffuse priors were used for most of the parameters, with certain exceptions.  For example, the prior for $\gamma_l$ ensures that the effect of medication level is protective.  Also, the parameters governing the medication level process $z_i$ were chosen to convey some information, guided by input from clinicians.  The prior for $\sigma$ and $\phi$ assumes a relatively consistent medication level in the absence of a true (discontinuous) change; allowing for the possibility of a change in medication level of approximately one step in six months.  Meanwhile $\rho$ was chosen to encourage a probability of anywhere from 0 to about a 25\% chance of suddenly changing medication in a given week.  Finally, the prior for $\varpi$ was chosen to reflect the perception that a patient is anywhere from 5\% to 75\%, with a best guess of about 33\%, likely to make a change in medication shortly after a ED/IP/OP/OB visit.

\begin{table}[t!]
\vspace{-.13in}
\begin{center}
\caption{Summary of the model for asthma related events defined in (\ref{eq:model_1})~$-$~(\ref{eq:model_5}) and the specification of prior distributions.}
\label{tab:priors}
\vspace{-.1in}
{%\small
  \renewcommand{\tabcolsep}{4pt}
\begin{tabular}{|l|c|c|c|}
  \hline
  Description & Model & Prior Distributions & Specification \\
  \hline
  \hline
  \multirow{4}{*}{Individual Effects} & \multirow{4}{*}{$\balpha_{i} \stackrel{iid}{\sim}  N(\bnu, \bSigma)$, $i=1,...,n$} & \multirow{2}{*}{$\nu_l \stackrel{ind}{\sim}  N(M_{\nu,l}, S^2_{\nu,l})$} & $M_{\nu,l}= 0$ \\
  &  & & $S^2_{\nu,l}= 100$ \\
  \cline{3-4}
  & & \multirow{2}{*}{$\bSigma \sim \mbox{Wishart}(\bP_\Sigma, N_\Sigma)$} & $\bP_\Sigma= \bI$ \\
  & & & $N_\Sigma=4$ \\
  \hline
  \hline
  \multirow{2}{*}{Fixed Effects} & \multirow{2}{*}{$\beta_{j,l}$ in (\ref{eq:model_2})} & \multirow{2}{*}{$\beta_{j,l} \stackrel{ind}{\sim}  N(B_{j,l}, T^2_{j,l})$} & $B_{j,l}= 0$ \\
  &  & & $T^2_{j,l}= 100$ \\
  \hline
  \hline
  \multirow{2}{*}{Medication Effects} & \multirow{2}{*}{$\gamma_{l}$ in (\ref{eq:model_2})} & \multirow{2}{*}{$-\gamma_{l} \stackrel{ind}{\sim}  \log N(C_{l}, U^2_{l})$} & $C_{l}= -1$ \\
  &  & & $U^2_{l}= 10$ \\
  \hline
  \hline
  \multirow{2}{*}{} & \multirow{2}{*}{$\bdelta_{i} {\sim}  GP(0, K)$, $i=1,...,n$} & \multirow{2}{*}{$e^{-\theta} \sim  \mbox{Beta}(A_\theta, B_\theta)$} & $A_\theta= 1$ \\
  Individual Time &  & & $B_\theta= 1$ \\
  \cline{3-4}
  Effect  & \multirow{2}{*}{$K$ in (\ref{eq:model_4})} & \multirow{2}{*}{$\bPhi \sim \mbox{Wishart}(\bP_\Phi, N_\Phi)$} & $\bP_\Phi= \bI$ \\
  & & & $N_\Phi=4$ \\
  \hline
  \hline
  \multirow{8}{*}{} & \multirow{10}{*}{$z_{i}$ in (\ref{eq:model_5})} & \multirow{2}{*}{$\mu_h \sim N(M_\mu, S^2_\mu)$} & $M_\mu=0$ \\
    &  & & $S^2_\mu= 50$ \\
  \cline{3-4}
    &  & \multirow{2}{*}{$\sigma^2 \sim \mbox{Gamma}^{-1}(A_\sigma, B_\sigma)$} & $A_\sigma= 5$ \\
  Medication & & & $B_\sigma=.05$ \\
  \cline{3-4}
  Level Process  &  & \multirow{2}{*}{$e^{-\phi} \sim  \mbox{Beta}(A_\phi, B_\phi)$} & $A_\phi= 6$ \\
  &  & & $B_\phi= 3$ \\
  \cline{3-4}
   &  & \multirow{2}{*}{$\rho \sim  \mbox{Gamma}(A_\rho, B_\rho)$} & $A_\rho= 0.25$ \\
  &  & & $B_\rho= 10$ \\
  \cline{3-4}
   &  & \multirow{2}{*}{$\varpi \sim  \mbox{Beta}(A_\varpi, B_\varpi)$} & $A_\varpi= 3$ \\
  &  & & $B_\varpi= 6$ \\
\hline
\end{tabular}
}
\end{center}
\vspace{-.0in}
\end{table}

Since fairly weak priors distributions were used for most parameters, there should be very little prior sensitivity with respect to such parameters.  However, parameters governing the medication level process $z_i$ had to be chosen more carefully since there is some confounding between the parameters $(\sigma, \phi)$ and $(\rho,\varpi)$.  That is, if the medication level needs to change a lot in a given window of time, it can do it either via a big variance for the continuous process or via many discrete jumps.  Thus, it is important to assess the extent of the sensitivity of the results to the choices for these priors.  Therefore, the MEPS analysis in Section~\ref{sec:MEPS} was conducted with prior parameter values specified in Table~\ref{tab:priors}, and also with values that conveyed less prior information.  Namely, values of $A_\sigma= 2$, $A_\sigma= .02$, $A_\phi=2$, $B_\phi=1$,  $A_\rho= 0.1$, $B_\rho= 4$, $A_\varpi= 3$, and $B_\varpi= 6$ were also used.  The parameter estimates (and thus also the predictions) did not change in any significant manner.  Finally, very diffuse values of $A_\sigma= .1$, $A_\sigma= .02$, $A_\phi=1$, $B_\phi=1$,  $A_\rho= 0.1$, $B_\rho= 1$, $A_\varpi= 1$, and $B_\varpi= 1$ were then used.  In this case, the estimates were qualitatively similar, but a bit different in magnitude (smaller) for the medication effect.  This is because with these very diffuse prior parameter settings, it favors a latent medication process that makes changes continuously as opposes to abruptly (i.e., unlike that in Figure~\ref{fig:z_i}.  This is a bit in contrast to our knowledge and intuition of medication patterns.  Thus, some care is required when specifying prior distributions for this model.  Since it can be challenging to translate knowledge about the medication process into the priors directly, a useful sanity check is to do the following.  Generate several curves from the proposed prior and make sure that the resulting latent medication process realizations that the prior can produce are consistent with what is known about the process.

The posterior distribution was approximated via MCMC.  The complete list of parameters sampled in the MCMC algorithm is
\bdm
\Theta = \left\{
\{\balpha_i\}_{i=1}^n,
\{\bdelta_i\}_{i=1}^n,
\{\tilde{z}_{i,1}, \dots, \tilde{z}_{i,H}\}_{i=1}^n,
\{B_i\}_{i=1}^n,
\{\bbeta_j\}_{j=1}^J,
\bgamma,
\bSigma,
\theta,
\bPsi,
\mu,
\sigma^2,
\phi,
\varpi
\right\},
\label{eq:param_list}
\edm
where $\bbeta_j=[\beta_{j,1},\beta_{j,2},\beta_{j,3}]'$, $\bgamma=[\gamma_{1},\gamma_{2},\gamma_{3}]'$, and the sequence of $\tilde{z}_{i,h}$ is truncated at $h=H$.  The upper bound $H$ can be chosen such that it is large enough not to affect the results.  A value of $H=10$ was used to generate the results below and was deemed sufficient since all of the posterior samples had $\max_{i,t} B_i(t) < 10$.  As was done in \citet*{Storlie12a}, the continuous functions $B_i$, $\tilde{z}_{i,h}$ and $\delta_{i,l}$ are resolved on a fine time grid, i.e., $B_{i} \equiv [B_{i}(t_{i,1}),\dots,B_{i}(t_{i,N_i})]'$ and similarly for the $\tilde{z}_{i,h}$ and $\delta_{i,l}$.  The corresponding integrals involving these functions needed to evaluate the likelihood are then approximated with quadrature.  One week granularity was used for the $t_{i,n}$ to match that used for the response in (\ref{eq:responses}).

The MCMC was carried out within a Gibbs framework via the \verb1rjags1 package in \verb1R1 \citep*{Plummer2009}, where each of the elements of $\Theta$ are updated in turn.  Five parallel chains were run for 20,000 iterations, the first 10,000 of which were discarded as burn-in.  All chains converged to the same distribution, as gauged by the non-individual specific (not $i$ dependent) parameters, and were thus aggregated.  On the MEPS dataset with $n=4,235$ individuals ($\sim300,000$ total time periods), these 20,000 iterations took $\sim40$ hours.  The Jags model \verb1R1 code is provided in a GitHub repository at \verb1https://github.com/cbstorlie/asthma_stepdown1.  Jags is a very convenient, general framework with which to conduct Gibbs sampling for Bayesian models, however, this convenience can come at the cost of computation time.  In this hierarchical model, for each MCMC iteration, the individual specific parameters $\{ \balpha_i, \bdelta_i, \tilde{z}_{i,1}, \dots, \tilde{z}_{i,H}, B_i \}$ could, in principle, be updated for each $i$ independently and thus in parallel.  This would substantially decrease computation time, however, this is not possible when using Jags and would require custom MCMC code.  Since training is done off-line and individual predictions are quite fast (on the order of $\sim$1 second), Jags seems sufficient for this application.

\section{MEPS Analysis Results}
\label{sec:MEPS}

\subsection{Inference}
The patient specific variables considered in addition to medication level were {\em sex}, {\em age}, residing in a metropolitan statistical area ({\em msa}), smoking status ({\em smoke1} = 1 for smoker, {\em smoke2} = 1 for non-smoker but with smoking exposure), Charlson sum of diseases ({\em ch\_count}), depression/anxiety ({\em dep}), chronic obstructive pulmonary disease ({\em copd}), gastroesophageal reflux disease ({\em gerd}), and rhinitis/sinusitis ({\em resp}).  These variables were chosen as they were the only variables readily available from MEPS that were thought to potentially have an effect on the outcomes. Other variables such as eosinophil levels, and pulmonary function, were not available.
The posterior distribution for the fixed effects is summarized in Table~\ref{tab:estimates}.  Increased controller medication level is strongly related to a lower rate of ED/IP visits and OCS usage.  However, medication level appears to have a much smaller influence on rescue inhaler usage.  Age is the only other factor among those investigated that has a significant effect.  According to the model, older patients are less likely to have an ED/IP visit, while they are more likely to use OCS.

\begin{table}[!h]
  \centering
  \caption{Posterior summary of effects for each of the three outcomes}
  \label{tab:estimates}
  \begin{tabular}{|l|rc|rc|rc|}
    \hline
    \multirow{2}{*}{Effect} & \multicolumn{2}{|c|}{ED/IP visit} & \multicolumn{2}{|c|}{Rescue Inhaler Usage} & \multicolumn{2}{|c|}{OCS Usage} \\
   & $\!$Estimate$\!$ & 95\% CI & $\!$Estimate$\!$ & 95\% CI & $\!$Estimate$\!$ & 95\% CI \\
    \hline
step\_level & $\! -0.264 \!$ & $\!\!( -0.388 , -0.183 )\!\!$ & $\! -0.017 \!$ & $\!\!( -0.023 , -0.013 )\!\!$ & $\! -0.189 \!$ & $\!\!( -0.228 , -0.161 )\!\!$\\
sex & $\!  0.270 \!$ & $\!\!( -0.307 ,  0.917 )\!\!$ & $\!  0.137 \!$ & $\!\!( -0.216 ,  0.426 )\!\!$ & $\!  0.082 \!$ & $\!\!( -0.350 ,  0.681 )\!\!$\\
age & $\! -0.028 \!$ & $\!\!( -0.046 , -0.013 )\!\!$ & $\! -0.009 \!$ & $\!\!( -0.016 ,  0.001 )\!\!$ & $\!  0.033 \!$ & $\!\!(  0.021 ,  0.042 )\!\!$\\
msa & $\! -0.294 \!$ & $\!\!( -0.941 ,  0.384 )\!\!$ & $\! -0.325 \!$ & $\!\!( -0.692 ,  0.032 )\!\!$ & $\!  0.318 \!$ & $\!\!( -0.154 ,  0.956 )\!\!$\\
smoke1 & $\! -0.244 \!$ & $\!\!( -1.035 ,  0.529 )\!\!$ & $\!  0.353 \!$ & $\!\!( -0.124 ,  0.697 )\!\!$ & $\! -0.029 \!$ & $\!\!( -0.605 ,  0.525 )\!\!$\\
smoke2 & $\! -0.033 \!$ & $\!\!( -0.794 ,  0.594 )\!\!$ & $\!  0.033 \!$ & $\!\!( -0.425 ,  0.408 )\!\!$ & $\! -0.106 \!$ & $\!\!( -0.917 ,  0.415 )\!\!$\\
ch\_count & $\! -0.412 \!$ & $\!\!( -0.941 ,  0.065 )\!\!$ & $\!  0.116 \!$ & $\!\!( -0.113 ,  0.323 )\!\!$ & $\!  0.008 \!$ & $\!\!( -0.279 ,  0.389 )\!\!$\\
dep\_flag2 & $\!  0.400 \!$ & $\!\!( -0.465 ,  1.285 )\!\!$ & $\!  0.108 \!$ & $\!\!( -0.400 ,  0.648 )\!\!$ & $\! -0.615 \!$ & $\!\!( -1.318 ,  0.067 )\!\!$\\
copd\_flag2 & $\!  0.298 \!$ & $\!\!( -0.318 ,  0.863 )\!\!$ & $\!  0.467 \!$ & $\!\!(  0.127 ,  0.931 )\!\!$ & $\! -0.295 \!$ & $\!\!( -0.759 ,  0.141 )\!\!$\\
gerd\_flag2 & $\! -0.204 \!$ & $\!\!( -0.946 ,  0.444 )\!\!$ & $\!  0.017 \!$ & $\!\!( -0.387 ,  0.426 )\!\!$ & $\!  0.226 \!$ & $\!\!( -0.290 ,  0.651 )\!\!$\\
resp\_flag2 & $\! -0.069 \!$ & $\!\!( -0.647 ,  0.449 )\!\!$ & $\!  0.051 \!$ & $\!\!( -0.321 ,  0.432 )\!\!$ & $\! -0.337 \!$ & $\!\!( -0.818 ,  0.275 )\!\!$\\
    \hline
  \end{tabular}
  \vspace{.15in}
\end{table}

%\clearpage
\vspace{-.25in}
\subsection{Prediction of an Adverse Event}

Let an adverse event for a six month period be defined as any ED/IP visit, any OCS use, or more than 3 rescue refills.  For a given draw of the parameters from the MCMC sample, the probability of an adverse event for a given patient at a given medication level can be obtained analytically based on the Poisson CDF and the {\em conditional} independence of the three outcomes.  The posterior predictive probability of an adverse event can then be obtained as the posterior mean of these probabilities.
Example patient 1 (14$\th$ individual in the data set), was at medication levels of 3, 2, 4, 5, 0 in his/her five rounds, respectively (see Table~\ref{tab:patient_sum}).  This patient had an adverse event in rounds 2 and 5 according to the definition above.  The posterior probability of an adverse event in the next 6 months if he/she stays at medication step level 0 is 0.453 (Table~\ref{tab:patient_pred}).  It may seem more appropriate for this patient to be on an at least a step level of 3 in order to lower this probability to a more reasonable $\sim$20\% level.

\renewcommand{\arraystretch}{1.1} 
\begin{table}[t!]
  \centering
  \caption{Summary data for patients 1 and 2 over the five rounds of the study.}
  \label{tab:patient_sum}
  \begin{tabular}{cc|ccccc|}
    \cline{3-7}
    & & \multicolumn{5}{|c|}{Round} \\
    & & 1 & 2 & 3 & 4 & 5 \\
    \hline
\multicolumn{1}{|c|}{\multirow{4}{*}{Patient 1}} & \multicolumn{1}{|c|}{Step Level} &  3 & 2 & 4 & 5 & 0 \\
\multicolumn{1}{|c|}{}& \multicolumn{1}{|c|}{ED/IP Visits}  & 0 & 0 & 0 & 0 & 2 \\
\multicolumn{1}{|c|}{}& \multicolumn{1}{|c|}{Rescue Inhaler} & 1 & 0 & 0 & 0 & 0 \\
\multicolumn{1}{|c|}{}& \multicolumn{1}{|c|}{OCS Use} & 0 & 1 & 0 & 0 & 0 \\
 \hline
\multicolumn{1}{|c|}{\multirow{4}{*}{Patient 2}} & \multicolumn{1}{|c|}{Step Level} &  3 & 3 & 0 & 3 & 3 \\
\multicolumn{1}{|c|}{}& \multicolumn{1}{|c|}{ED/IP Visits}  & 0 & 0 & 0 & 0 & 0 \\
\multicolumn{1}{|c|}{}& \multicolumn{1}{|c|}{Rescue Inhaler} & 0 & 0 & 0 & 0 & 0 \\
\multicolumn{1}{|c|}{}& \multicolumn{1}{|c|}{OCS Use} & 0 & 0 & 0 & 0 & 0 \\
 \hline
  \end{tabular}
\end{table}

\renewcommand{\arraystretch}{1.1} 
\begin{table}[h!]
  \centering
  \caption{Posterior probability of adverse event in next 6 months for patients 1 and 2 at each of the possible medication step levels.  The medication level the respective patient was at in the last round of the study is in bold.}
  \label{tab:patient_pred}
\vspace{-.1in}
  \begin{tabular}{c|c|c|}
    \cline{2-3}
    & \multirow{2}{*}{Step Level}  & Probability of \\
    &  & Adverse Event \\
\hline
\multicolumn{1}{|c|}{\multirow{6}{*}{Patient 1}}  & \bf 0 & \bf 0.453 \\
\multicolumn{1}{|c|}{}    & 1 & 0.369 \\
\multicolumn{1}{|c|}{}    & 2 & 0.291 \\
\multicolumn{1}{|c|}{}    & 3 & 0.226 \\
\multicolumn{1}{|c|}{}    & 4 & 0.175 \\
\multicolumn{1}{|c|}{}    & 5 & 0.131 \\
    \hline
\multicolumn{1}{|c|}{\multirow{6}{*}{Patient 2}}  &  0 &  0.142 \\
\multicolumn{1}{|c|}{}    & 1 & 0.111 \\
\multicolumn{1}{|c|}{}    & 2 & 0.086 \\
\multicolumn{1}{|c|}{}    & \bf 3 & \bf 0.067 \\
\multicolumn{1}{|c|}{}    & 4 & 0.051 \\
\multicolumn{1}{|c|}{}    & 5 & 0.040 \\
    \hline
   \end{tabular}
\end{table}

Conversely, patient 2 (24$\th$ individual in the data set) was at step level 3 in the last round.  The probability of an adverse event at each of the medication step levels for the next six month period is also provided in Table~\ref{tab:patient_pred}.  If this patient remains at step level 3, the probability of an adverse event is quite low at 0.067.  The provider and patient may decide that it is appropriate to reduce the patient's medication level in this case.

\vspace{.1in}
\subsection{Validation}

Validating the predictive accuracy of the model using MEPS data is difficult due to the fact that medication level over time is not entirely known within a round.  However, just for this validation exercise, it was assumed that medication level remained constant during the last round (round 5) for a set of 500 patients.  All observations from round 5 for 500 randomly selected patients were held out of the model fitting process and the model was then used to predict the probability of an adverse event during round 5 for each of these patients.
This validation exercise is conservative since the model was given potentially false information for medication level (assumed constant) as input, yet it has to predict the reality (produced by medication level changing during a round).  Even still, comparison of such predictions top the observed outcomes should provide a useful sanity check.  

\begin{figure}[t!]
\vspace{-.0in}
  \centering
  \caption{(a) Histogram of predicted probabilities for an adverse event in round five for the 500 held out observations. (b) Calibration plot of the predicted probability versus the observed probability (aggregated within different strata of risk).}
    \label{fig:calibration}
    \begin{subfigure}[b]{.49\textwidth}
      \centering
\vspace{-.1in}
      \caption{}
\vspace{-.0in}
      \includegraphics[width=.82\textwidth]{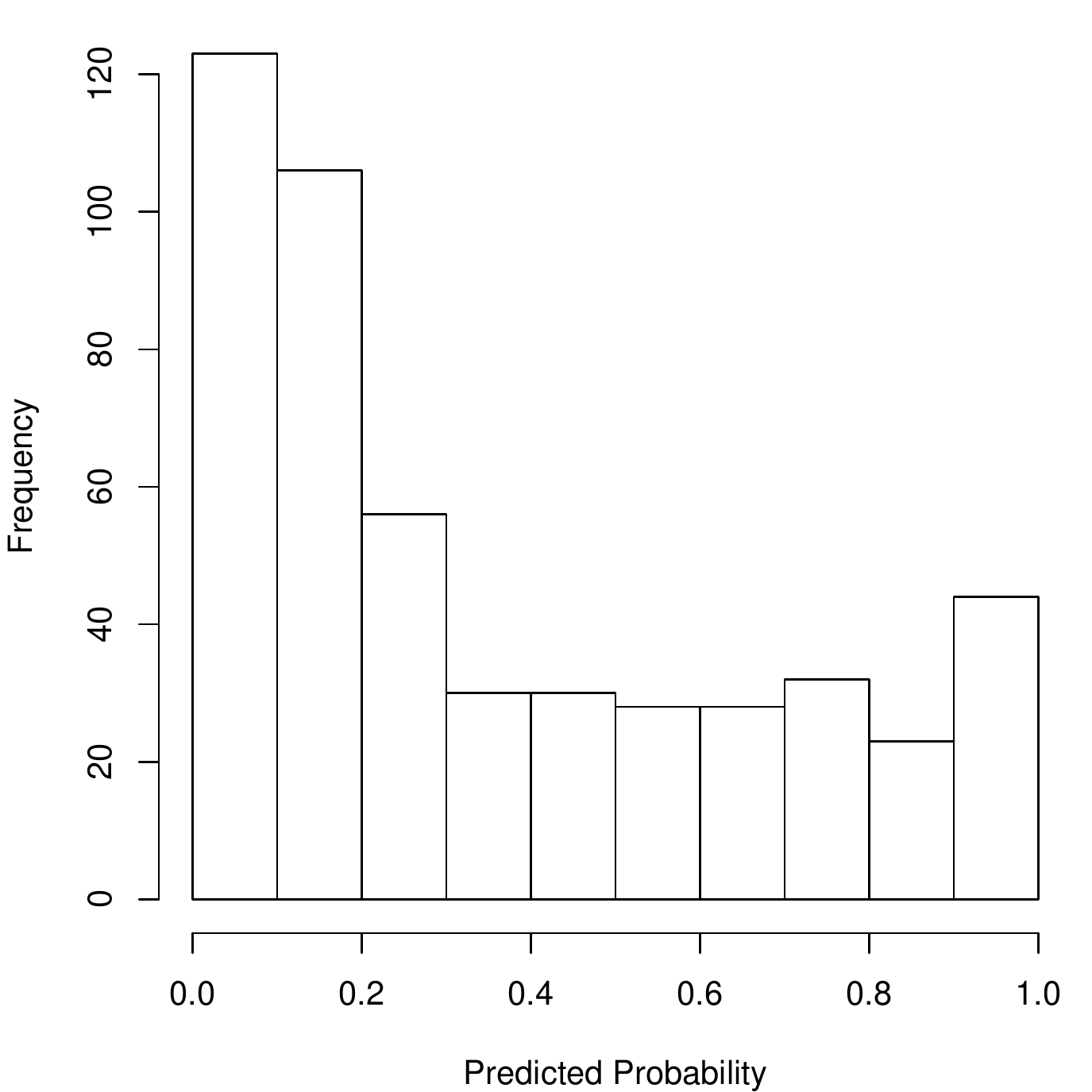}
    \end{subfigure}
    \begin{subfigure}[b]{.49\textwidth}
      \centering
\vspace{-.1in}
      \caption{}
\vspace{-.3in}
      \includegraphics[width=.92\textwidth]{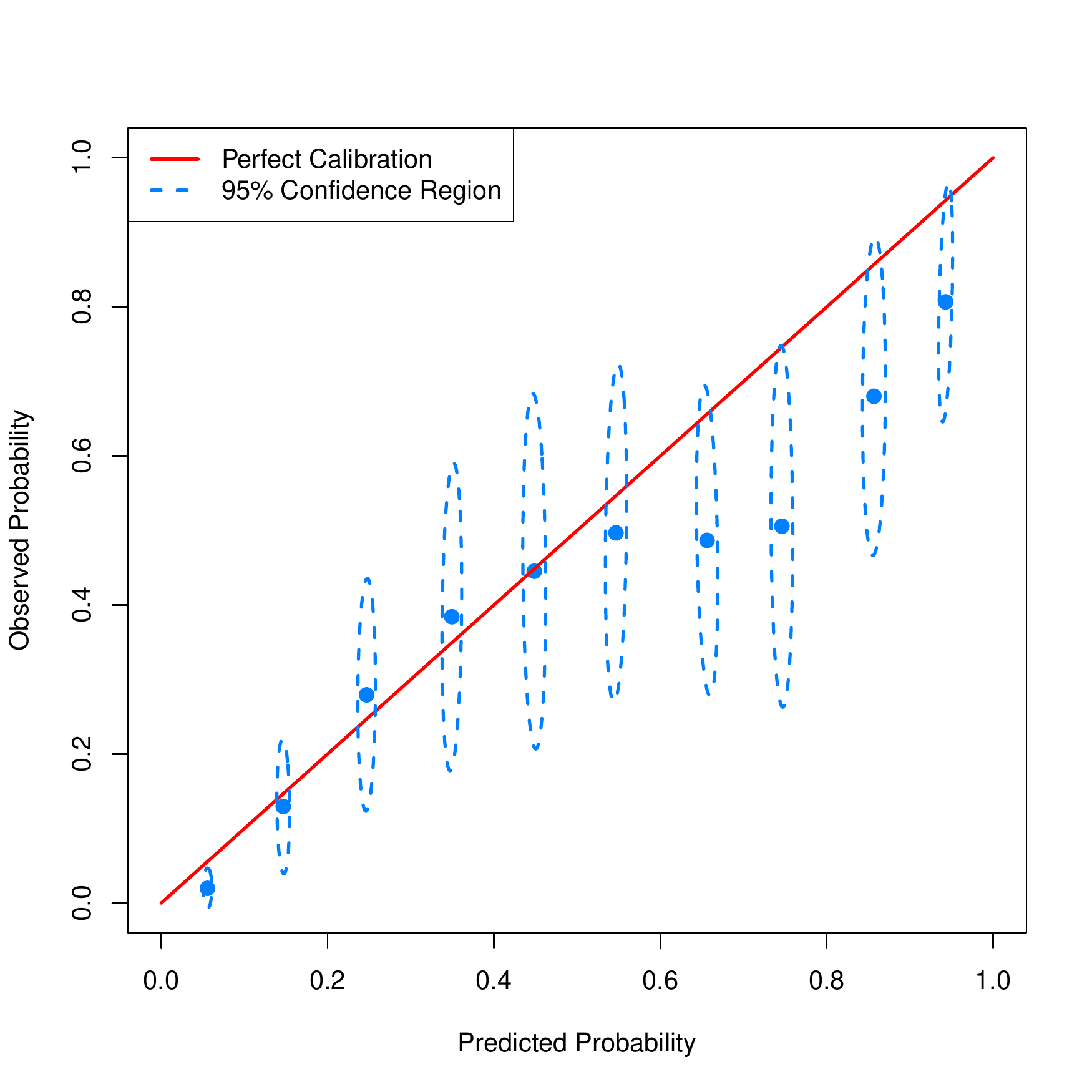}
    \end{subfigure}
    \begin{subfigure}[b]{.49\textwidth}
      \centering
\vspace{.1in}
      \caption{}
\vspace{-.3in}
      \includegraphics[width=.92\textwidth]{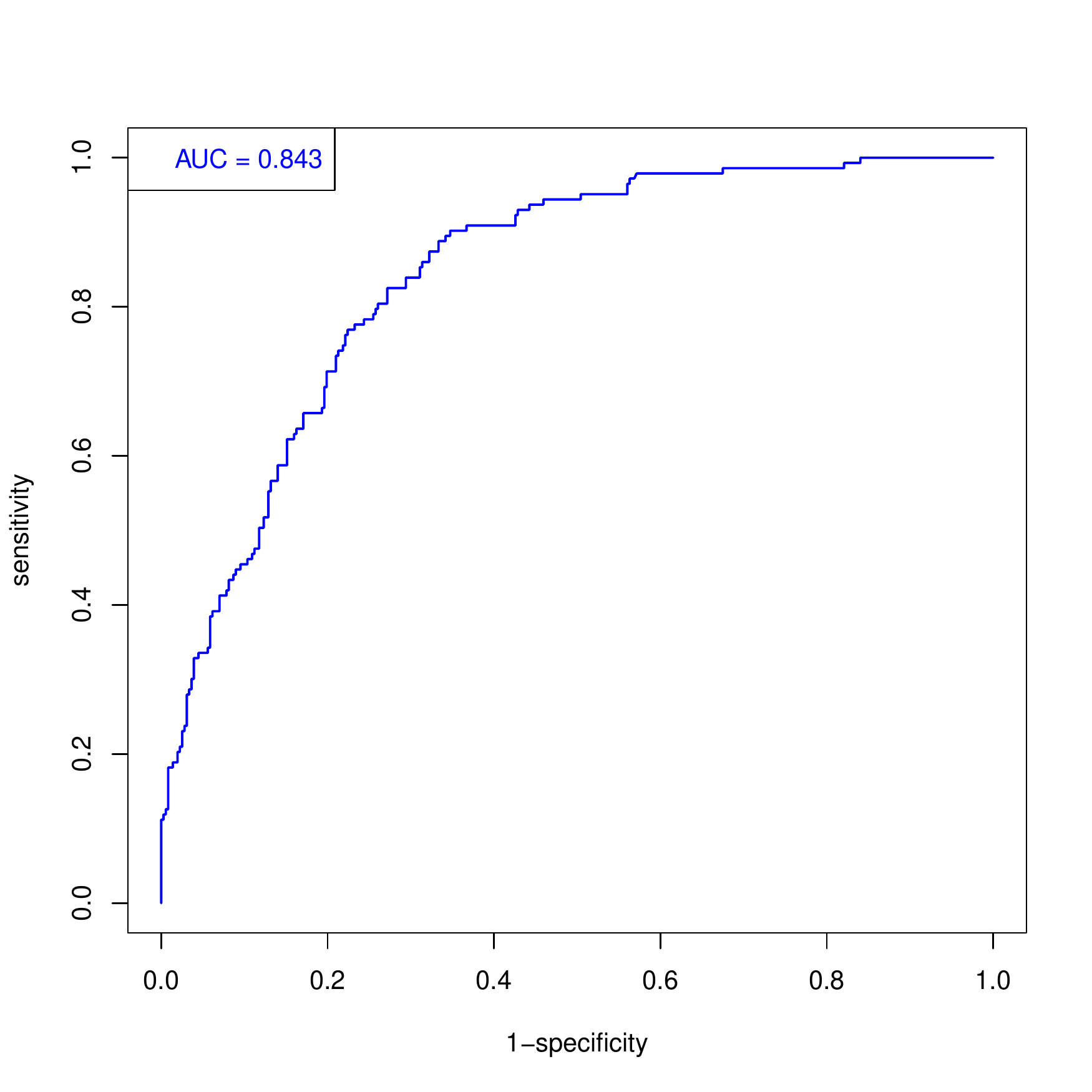}
    \end{subfigure}
\vspace{-.0in}
\end{figure}

The histogram of the distribution of these predicted probabilities is provided in Figure~\ref{fig:calibration}(a).  The majority of predicted probabilities are below 0.25, but there is a relatively uniform spread of probabilities from 0.25 to 1.0.  Figure~\ref{fig:calibration}(b) provides a graphical comparison of the 500 predictions to the observed outcomes similar to that used in \citet*{Harrell96} and \citet*{Austin14}.  Since the outcomes for each individual are 0 or 1, they are aggregated into various bins of predicted risk (in this case the 10 bins resulting from the cut points $0.0, 0.1, \dots, 0.9, 1.0$).  The ten plotted points are the average of the predicted probability within the respective bin against the empirical (or observed) probability of an event within that bin.  If the model is reasonable, it would be expected that the blue points would fall close to the red $y=x$ line.
Individual 95\% confidence regions for the {\em predicted} versus {\em observed} points in each bin were obtained via a nonparametric bootstrap approach.  It would be expected that each of these intervals would intersect the red line 95\% of the time if the model was correct.

Intuition would suggest that the predictions provided by the model are almost certainly incorrect in this case, since the medication level being provided for prediction is likely not constant during the final round for many of the 500 patients.
Nonetheless, the model predictions appear to agree very well with the observations for smaller values of predicted risk.  There may be some upward bias in predictions of higher risk individuals, i.e., the model predictions are a bit conservative.  However, it is far less critical to have predictive probabilities above 0.5 align well with reality (i.e., anything approaching 0.5 is ``high'', and it is not important to distinguish a 0.6 from a 0.7, for example).  It is precisely the smaller risk predictions that need to be accurate for the clinical use case.
%That is, suppose the model predicts a risk of 0.05 for an adverse event in the next 6 months with a reduced medication level for a patient.  It is then important that the actual probability be very close to 0.05, and not really 0.15, for example.  If the model predicts a risk of 0.8 and the actual probability is 0.6, then this is inconsequential, since a medication step down is not advisable anyhow.
Fortunately, there is no evidence of prediction bias among the lower risk ($\leq0.5$) predictions.

\section{Simulation Study}
\label{sec:sims}

In this section, a further examination of the model performance is provided on a simulation case designed to be similar to the MEPS data.  Specifically, the posterior mean from the MEPS analysis for all parameters was used to generate a reduced data set for the first $n=500$ individuals.  The medication level for each patient was set to the average step level he/she had in round 1 in the MEPS data.  In order to mimic a key feature of the MEPS data, upon an ED/IP visit (as randomly generated by the model) it was assumed that, with probability 0.5, the medication step level would be {\em increased} to a randomly selected higher level (as opposed to a random change in either direction).  The same mechanism is applied to an OCS prescription fill as well.  There is also 0.3\% chance (the posterior mean estimate) that the medication level will change for a simulated patient on any given day (either up or down) in a random fashion with probability of the new level proportional to the frequency that step level was observed in the MEPS data.  This process was repeated to create a total of 100 synthetic data sets.

\begin{figure}[t!]
      \begin{widepage}  
  \vspace{-.1in}
  \begin{centering}
  \caption{Estimates from the proposed latent process approach and a simpler GLMER model obtained from the 100 simulated data sets. (a) Boxplots of estimates for $y_1\::\:$ED/IP Visit parameters. (b) Boxplots of estimates for $y_2\::\:$Rescue Inhaler Refills parameters. (c) Boxplots of estimates for $y_3\::\:$Use of OCS parameters.}
  \vspace{-.25in}
    \label{fig:sims_box}
    \begin{subfigure}[b]{.373\textwidth}
      \caption{ED/IP Visit}
      \includegraphics[width=.999\textwidth]{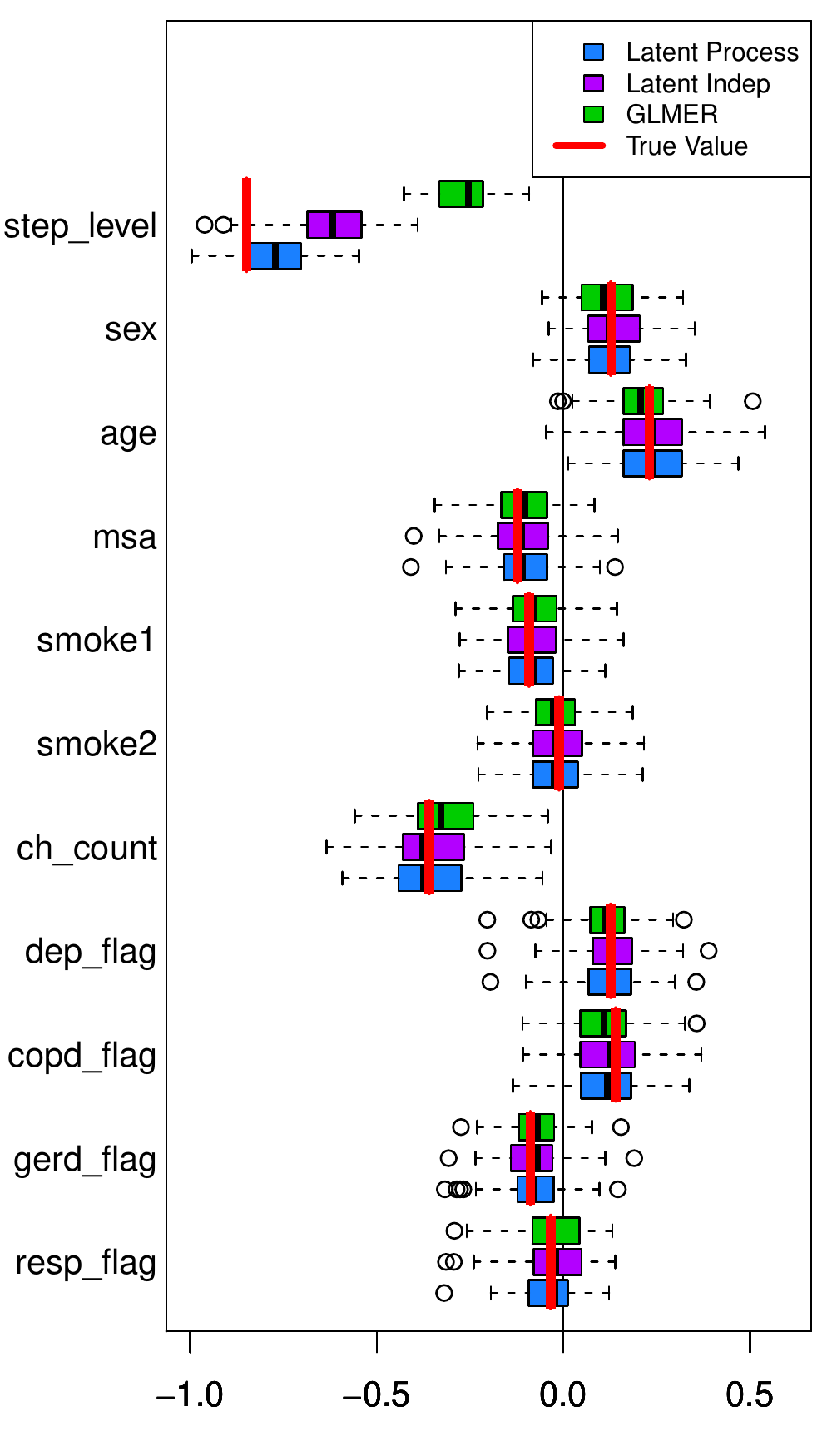}
    \end{subfigure}
    \begin{subfigure}[b]{.307\textwidth}
      \caption{Rescue Inhaler Refills$\;\;\;$}
      \includegraphics[width=.999\textwidth]{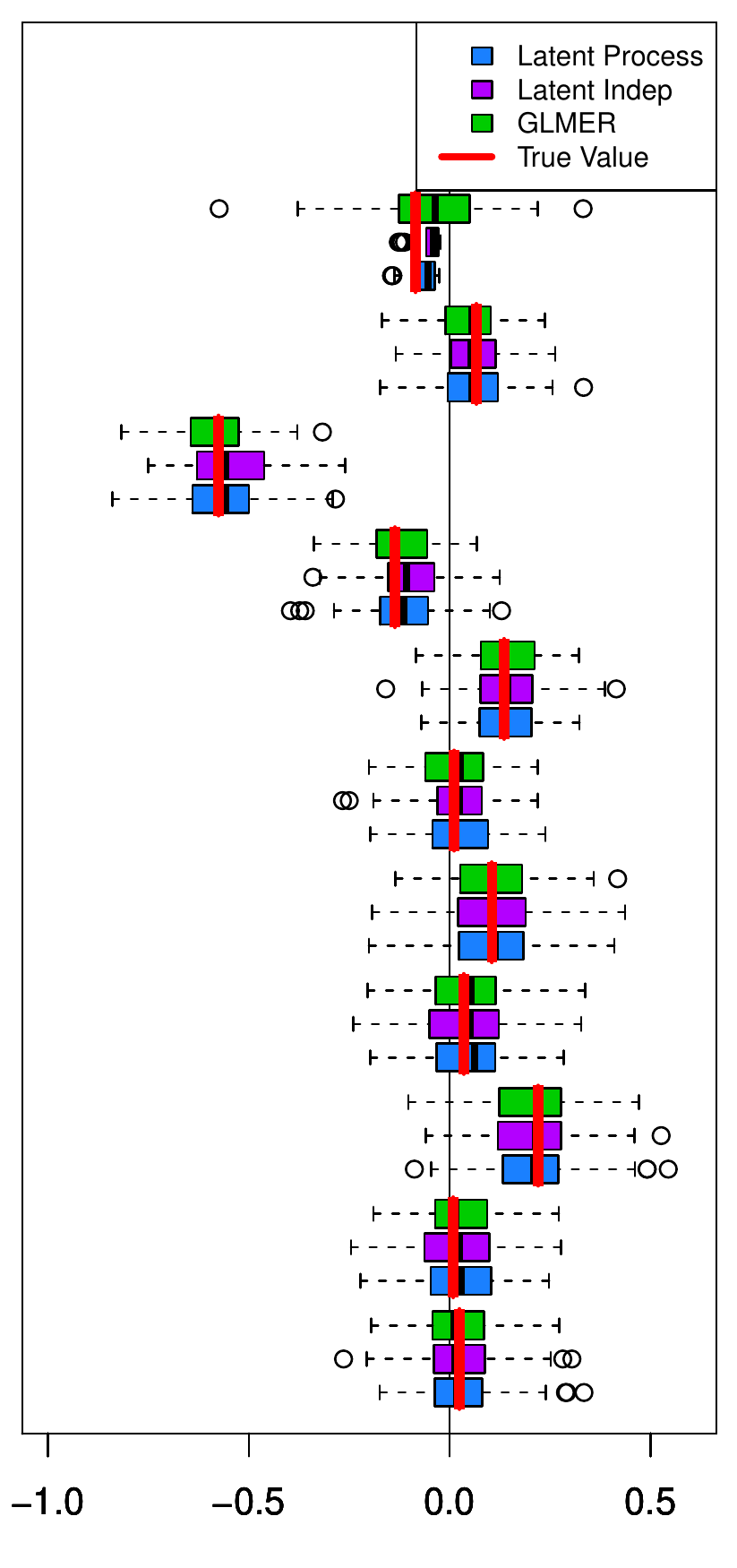}
    \end{subfigure}
    \begin{subfigure}[b]{.307\textwidth}
      \caption{OCS Use$\!$}
      \includegraphics[width=.999\textwidth]{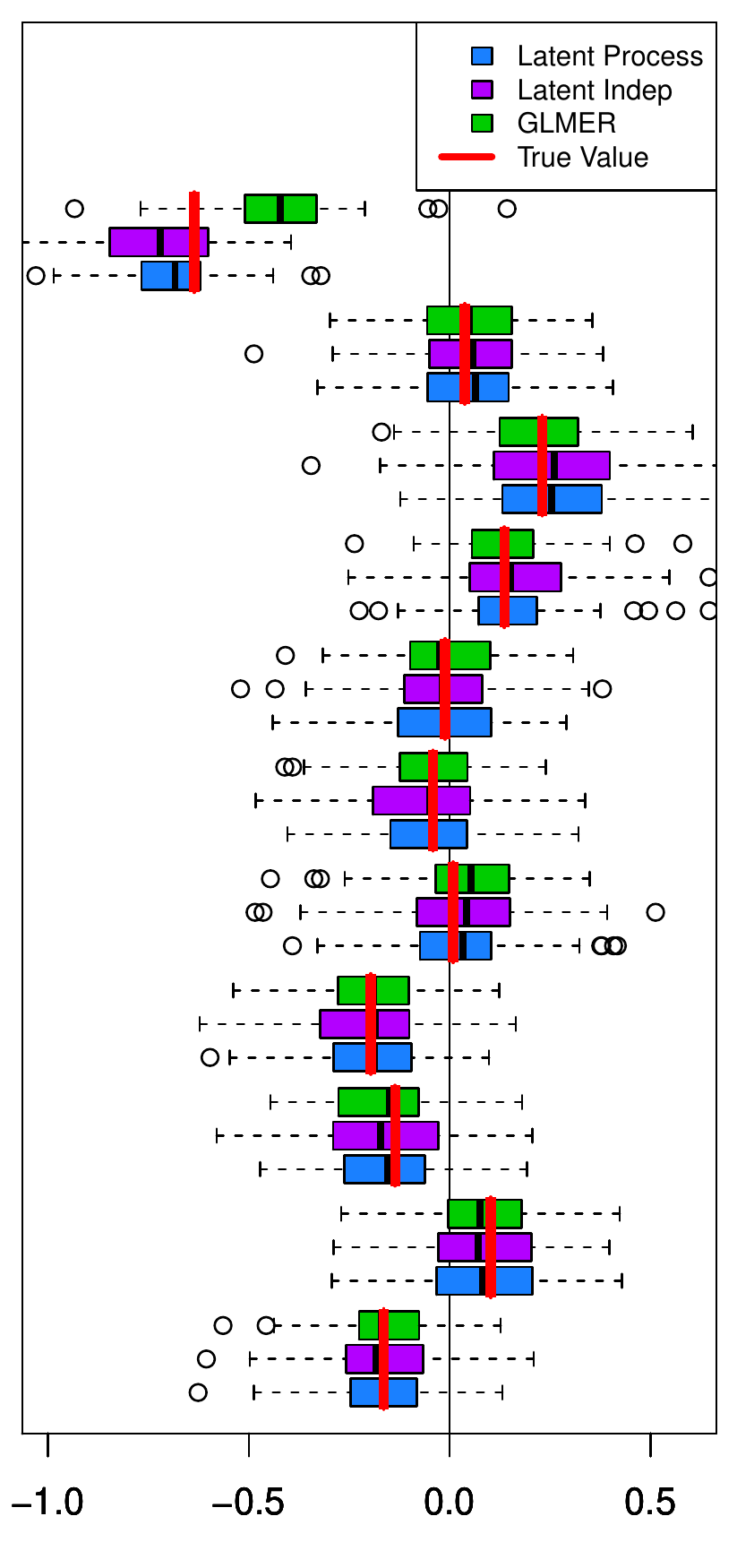}
    \end{subfigure}
  \end{centering}
      \end{widepage}
  \vspace{-.3in}
\end{figure}

The proposed model (Latent Process) was then fit to each of the 100 data sets and the resulting posterior means of the parameters in Table~\ref{tab:estimates} are summarized as boxplots in Figures~\ref{fig:sims_box}(a)~-~\ref{fig:sims_box}(c).  The ``observed'' data used to fit the model was kept consistent with the MEPS data.
That is, it was assumed that the {\em average} medication level for a round was the only information about controller medication available, ED/IP visits were known to the day, but rescue inhaler fills and OCS were only known in aggregate for an entire round.
The estimates provided in Figures~\ref{fig:sims_box} are the respective posterior means for each of the 100 data sets.  The coefficients have been standardized by the standard deviation of the respective variables to make them comparable.  A variant of the proposed approach was also used that assumed independent outcomes (Latent Indep), i.e., restricted such that $\bSigma$ and $\Psi$ are diagonal with independent inverse-gamma priors on the diagonal elements.  This approach was included to assess how much impact the treatment of multivariate dependence may have on the modeling results. A generalized linear mixed effects regression (GLMER) using the \verb1R1 package \verb1lme41 was also fit separately to each of the three responses.  The GLMER model has a random intercept for each patient like the proposed model, however, the medication level was assumed fixed at the average value observed for the round.  This was intended to provide an evaluation of what might be gained by treating the medication level as a latent process.  Boxplots of the corresponding GLMER estimates are provided in Figures~\ref{fig:sims_box}(a)~-~\ref{fig:sims_box}(c) as well.  The {\em true} value of each parameter used to generate the 100 datasets is provided as a red line in each boxplot for comparison.

\begin{figure}[t!]
\vspace{-.0in}
  \centering
  \caption{(a) Calibration plot of the predicted probability versus the true probability of an adverse event (i.e., either an ED/IP visit, $\geq 4$ rescue inhaler fills, or any OCS use) in the next 6 months across all 100 simulations (aggregated within different strata of risk). (b) ROC plot using predicted probability to predict observed events (generated randomly according to the true probabilities).}
    \label{fig:calibration_sim}
    \begin{subfigure}[b]{.49\textwidth}
      \centering
\vspace{.0in}
      \caption{}
\vspace{-.2in}
      \includegraphics[width=.99\textwidth]{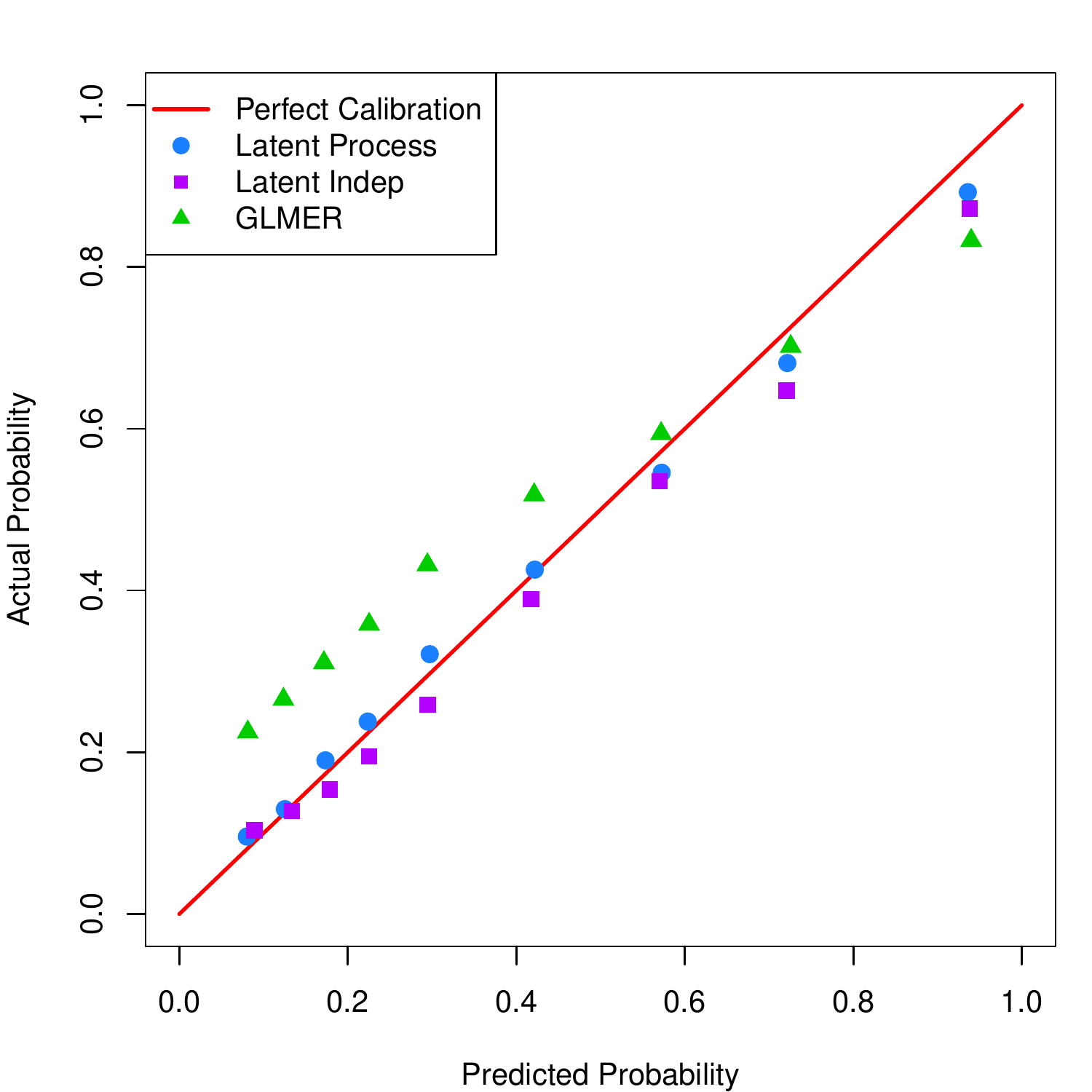}
    \end{subfigure}
    \begin{subfigure}[b]{.49\textwidth}
      \centering
\vspace{.0in}
      \caption{}
\vspace{-.2in}
      \includegraphics[width=.99\textwidth]{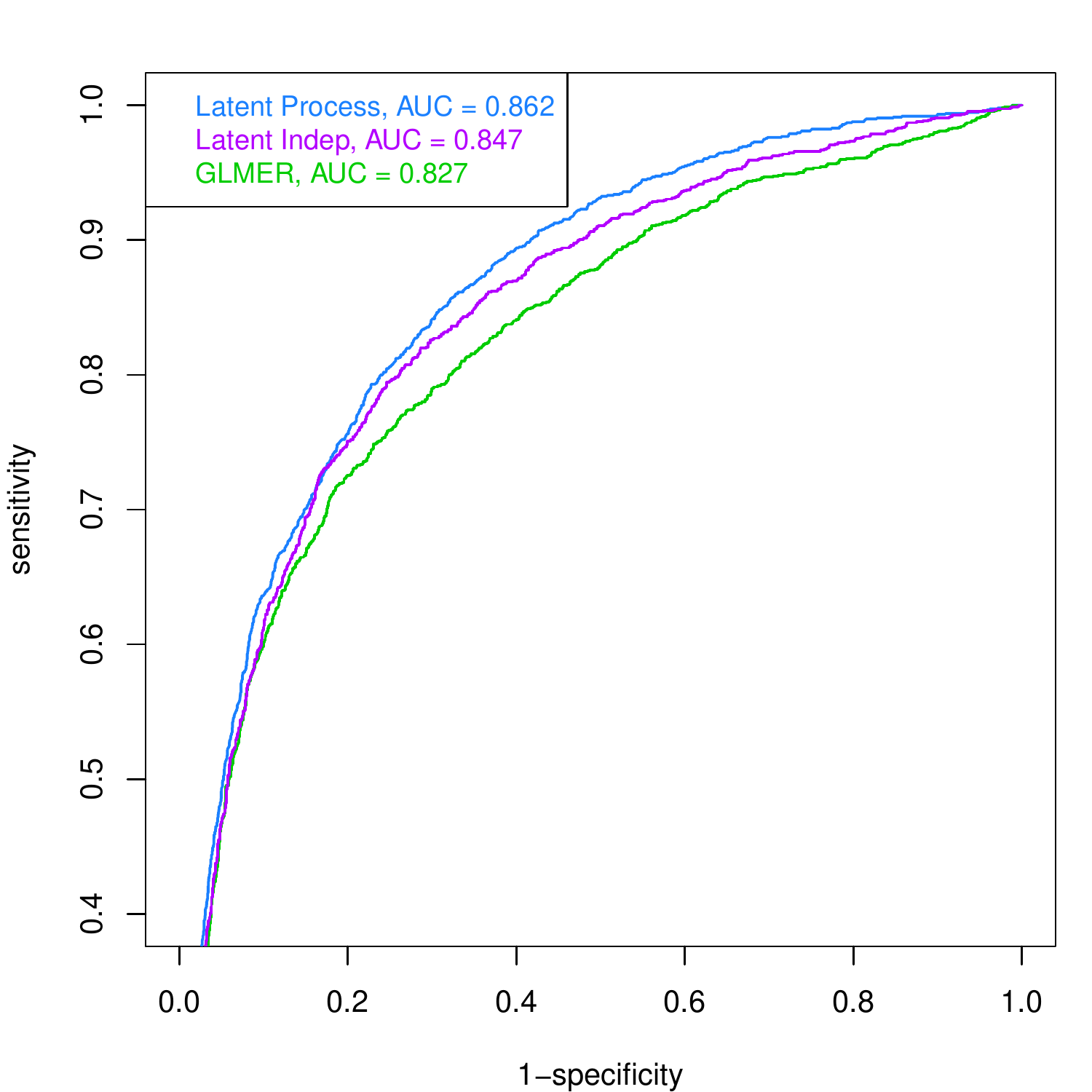}
    \end{subfigure}
\vspace{-.0in}
\end{figure}

It is clear from Figure~\ref{fig:sims_box}(a) that GLMER substantially underestimates the magnitude of the effect that medication has on ED/IP visits.  The proposed approach on the other hand, has a slight bias for this coefficient (as to be expected with a Bayesian approach) but it provides a far better estimate in general since it allows for the possibility of changes in medication to occur at/after an ED/IP visit.  The proposed latent medication approach with independent outcomes has less bias than GLMER, but it is not as close to the true value as the full multivariate approach.  The GLMER estimate is also biased toward zero on the effect that medication has on OCS use.  The practical ramifications of the substantial bias introduced by GLMER for this problem can be seen in Figure~\ref{fig:calibration_sim}.

Figure~\ref{fig:calibration_sim}(a) is a calibration plot similar to that in Figure~\ref{fig:calibration}(b), but with effectively no error in the estimated points due to the 100 repeated simulations providing enough data to make the area of the confidence regions negligible.
Also, the medication level for the patients to be predicted in Figure~\ref{fig:calibration_sim}(a) was randomly chosen to be a step {\em down} from the patients' respective medication level at the end of the simulated training data.  Restricting to a step down was done to highlight the issues that are present with GLMER predictions on the most important use case.  If a simulated patient was at step 0 at the end of the simulated training data, then they could not be stepped down, and were thus not included in this particular prediction exercise (usually about 150/500 were at step level 0 and excluded).  Figure~\ref{fig:calibration_sim}(a) shows that the predictions from the proposed approach are well calibrated, but that the GLMER predictions are substantially optimistic; an average probability of 0.07 according to GLMER leads to an actual probability of adverse event of over 0.20.  This would cause recommendations for medication step down to individuals that should not be lowering their medication level more often than is desirable.

Figure~\ref{fig:calibration_sim}(b) provides ROC curves for predicted probability of an event for the three approaches, using a randomly selected medication level for the next six months for each hypothetical subject. The figure demonstrates better overall predictive capability of the proposed approach, above and beyond better calibration for the case of a step down.  Table~\ref{tab:sim_Rsq} provides the $R^2$ measures of the predicted probability of an event in the next six months to the true probability of event for each outcome separately, and for the aggregated adverse event outcome (any).  This table demonstrates that the estimated probabilities obtained from the proposed latent process approach are far more accurate than those from GLMER which assumes that medication level is constant between rounds in addition to independent outcomes.  The {\em Latent Process} approach which is the full multivariate version of the proposed approach is also significantly more accurate than {\em Latent Indep} which still uses the latent process for medication, but assumes that the outcomes are dependent.  Interestingly, the {\em Latent Process} approach performs better prediction of the individual outcomes than {\em Latent Indep} due to the ability to borrow strength from each outcome when estimating an individual's overall and current level of risk for a particular outcome.

\begin{table}[!t]
  \centering
  \caption{$R^2$ of the predicted probability of an event in the next six months to the true probability of event; averaged over the 100 simulations along with (standard error) for each outcome: (i) any ED/IP visit, (ii) four or more rescue inhaler refills, (iii) any oral steroid use, (iv) exacerbation, i.e., any of (i), (ii), or (iii).}
  \label{tab:sim_Rsq}
  \begin{tabular}{|lcccc|}
    \hline
    \multirow{2}{*}{Method} & \multicolumn{4}{c|}{Outcome} \\
    & ED/IP & OCS & $\geq 4$ rescue & Any \\
    \hline
Latent Process & 0.713 (0.005) & 0.836 (0.002) & 0.765 (0.004) & 0.811 (0.002)\\
Latent Indep & 0.607 (0.005) & 0.818 (0.003) & 0.728 (0.004) & 0.779 (0.002)\\
GLMER & 0.454 (0.008) & 0.696 (0.006) & 0.657 (0.007) & 0.698 (0.005)\\
    \hline
  \end{tabular}
  \vspace{.15in}
\end{table}

\section{Conclusions \& Further Work}
\label{sec:conclusions}

In this paper we developed a general approach to prediction of individual outcomes for patients using prescription data that is ambiguous about the exact medication level a patient is at over time.  The statistical approach treated medication level as an uncertain latent process in a measurement error framework, to reduce the bias in outcome prediction.  This approach was applied to a population of asthma sufferers using the data available from the MEPS database years 2000-2010.  The analysis demonstrated that well-calibrated predictions can be made for the individual-specific probability of an adverse outcome in a future time period.  While there is some upward bias of predicted risk for larger risk individuals, the smaller risk predictions agreed very well with the validation set observations. The approach was also compared via a simulation study to a more standard GLMER model with a random intercept, ignoring the uncertainty in medication level.  These results showed the potential for GLMER to be significantly biased when predicting outcomes due to the incorrect assumption of constant medication level when training the model.

It would be prudent to validate the model prediction using a separate data set where the prescription refill dates are more precisely known.  Such data would also allow for improved model estimation, though the medication level would still need a latent process treatment due to the fact that refill dates do not provide information about  daily use, when medication is discontinued, etc.  The goal is to ultimately perform pilot trials of embedding the clinical prediction model within the healthcare delivery system using (i) shared-decision making tools used within the patient-provider visit, (ii) electronic medical record alerts directed to providers, and (iii) risk information shared directly with patients using secure messaging. 

%As with any statistical model, we have some assessment of validity of assumptions, which may result in changes, etc.  But once we settle in on a reasonable model and fit it, we can use the model to simulate a future reality for the patients and use an approach like that above to demonstrate that it does what it was designed to do (i.e., limit the probability of poor symptom control to the specified tolerance).  We could also hold out the last time period of the real data from the model fitting and use that last time period as a validation.

{%\small
%  \singlespacing
\bibliography{curt_ref.bib}

\begin{thebibliography}{99}

\bibitem[Austin and Steyerberg(2014)Austin and Steyerberg]{Austin14}
\textsc{Austin, Peter~C and Steyerberg, Ewout~W}. (2014).
\newblock Graphical assessment of internal and external calibration of logistic
  regression models by using loess smoothers.
\newblock {\em Statistics in medicine\/}~\textbf{33}(3), 517--535.

\bibitem[Bateman \emph{and others}(2015)Bateman, Buhl, O'Byrne, Humbert,
  Reddel, Sears, Jenkins, Harrison, Quirce, Peterson  et~al.]{Bateman2015}
\textsc{Bateman, Eric~D, Buhl, Roland, O'Byrne, Paul~M, Humbert, Marc, Reddel,
  Helen~K, Sears, Malcolm~R, Jenkins, Christine, Harrison, Tim~W, Quirce,
  Santiago, Peterson, Stefan  \emph{and others}}. (2015).
\newblock Development and validation of a novel risk score for asthma
  exacerbations: the risk score for exacerbations.
\newblock {\em Journal of Allergy and Clinical Immunology\/}~\textbf{135}(6),
  1457--1464.

\bibitem[Blakey \emph{and others}(2012)Blakey, Woulnough, James, Fellows,
  Obeidat, Navaratnam, Stringfellow, Yeoh, Pavord, Thomas  et~al.]{Blakey2012}
\textsc{Blakey, JD, Woulnough, K, James, AC, Fellows, J, Obeidat, M,
  Navaratnam, V, Stringfellow, T, Yeoh, ZW, Pavord, I, Thomas, M  \emph{and
  others}}. (2012).
\newblock S62 a systematic review of factors associated with future asthma
  attacks to inform a risk assessment questionnaire.
\newblock {\em Thorax\/}~\textbf{67}(Suppl 2), A31--A32.

\bibitem[Brix and Diggle(2001)Brix and Diggle]{brix2001spatiotemporal}
\textsc{Brix, Anders and Diggle, Peter~J}. (2001).
\newblock Spatiotemporal prediction for log-gaussian cox processes.
\newblock {\em Journal of the Royal Statistical Society: Series B (Statistical
  Methodology)\/}~\textbf{63}(4), 823--841.

\bibitem[Carroll \emph{and others}(2006)Carroll, Ruppert, Stefanski and
  Crainiceanu]{carroll2006}
\textsc{Carroll, Raymond~J, Ruppert, David, Stefanski, Leonard~A and
  Crainiceanu, Ciprian~M}. (2006).
\newblock {\em Measurement error in nonlinear models: a modern perspective\/},
  2nd edition. Boca Raton, FL: CRC press.

\bibitem[Dunson(2000)Dunson]{Dunson2000}
\textsc{Dunson, David~B}. (2000).
\newblock Bayesian latent variable models for clustered mixed outcomes.
\newblock {\em Journal of the Royal Statistical Society: Series B (Statistical
  Methodology)\/}~\textbf{62}(2), 355--366.

\bibitem[Harrell \emph{and others}(1996)Harrell, Lee and Mark]{Harrell96}
\textsc{Harrell, Frank~E, Lee, Kerry~L and Mark, Daniel~B}. (1996).
\newblock Tutorial in biostatistics multivariable prognostic models: issues in
  developing models, evaluating assumptions and adequacy, and measuring and
  reducing errors.
\newblock {\em Statistics in medicine\/}~\textbf{15}, 361--387.

\bibitem[Jang \emph{and others}(2013)Jang, Chan, Huang and Sullivan]{Jang2013}
\textsc{Jang, Junho, Chan, Kwun Chuen~Gary, Huang, Hsiang and Sullivan,
  Sean~D}. (2013).
\newblock Trends in cost and outcomes among adult and pediatric patients with
  asthma: 2000--2009.
\newblock {\em Annals of Allergy, Asthma \& Immunology\/}~\textbf{111}(6),
  516--522.

\bibitem[Kamble and Bharmal(2009)Kamble and Bharmal]{Kamble2009}
\textsc{Kamble, Shital and Bharmal, Murtuza}. (2009).
\newblock Incremental direct expenditure of treating asthma in the united
  states.
\newblock {\em Journal of Asthma\/}~\textbf{46}(1), 73--80.

\bibitem[M{\o}ller \emph{and others}(1998)M{\o}ller, Syversveen and
  Waagepetersen]{moller1998log}
\textsc{M{\o}ller, Jesper, Syversveen, Anne~Randi and Waagepetersen,
  Rasmus~Plenge}. (1998).
\newblock Log gaussian cox processes.
\newblock {\em Scandinavian journal of statistics\/}~\textbf{25}(3), 451--482.

\bibitem[Moorman \emph{and others}(2012)Moorman, Akinbami, Bailey, Zahran,
  King, Johnson and Liu]{Moorman2012}
\textsc{Moorman, Jeanne~E, Akinbami, Lara~J, Bailey, CM, Zahran, HS, King, ME,
  Johnson, CA and Liu, X}. (2012).
\newblock National surveillance of asthma: United states, 2001-2010.
\newblock {\em Vital \& health statistics. Series 3, Analytical and
  epidemiological studies/[US Dept. of Health and Human Services, Public Health
  Service, National Center for Health Statistics]\/}~\textbf{35}, 1--58.

\bibitem[Muthen(1983)Muthen]{Muthen83}
\textsc{Muthen, Bengt}. (1983).
\newblock Latent variable structural equation modeling with categorical data.
\newblock {\em Journal of Econometrics\/}~\textbf{22}(1), 43--65.

\bibitem[Plummer(2009)Plummer]{Plummer2009}
\textsc{Plummer, M}. (2009).
\newblock rjags: Interface to the jags mcmc library.
\newblock {\em R package version\/}~\textbf{1}(3).

\bibitem[Rank \emph{and others}(2016)Rank, Liesinger, Branda, Gionfriddo,
  Schatz, Zeiger and Shah]{Rank2016}
\textsc{Rank, Matthew~A, Liesinger, Juliette~T, Branda, Megan~E, Gionfriddo,
  Michael~R, Schatz, Michael, Zeiger, Robert~S and Shah, Nilay~D}. (2016).
\newblock Comparative safety and costs of stepping down asthma medications in
  patients with controlled asthma.
\newblock {\em Journal of Allergy and Clinical Immunology\/}~\textbf{137}(5),
  1373--1379.

\bibitem[Rank \emph{and others}(2012)Rank, Liesinger, Ziegenfuss, Branda, Lim,
  Yawn, Li and Shah]{Rank2012a}
\textsc{Rank, Matthew~A, Liesinger, Juliette~T, Ziegenfuss, Jeanette~Y, Branda,
  Megan~E, Lim, Kaiser~G, Yawn, Barbara~P, Li, James~T and Shah, Nilay~D}.
  (2012{\em a}).
\newblock Asthma expenditures in the united states comparing 2004 to 2006 and
  1996 to 1998.
\newblock {\em The American journal of managed care\/}~\textbf{18}(9),
  499--504.

\bibitem[Rank \emph{and others}(2012)Rank, Liesinger, Ziegenfuss, Branda, Lim,
  Yawn and Shah]{Rank2012b}
\textsc{Rank, Matthew~A, Liesinger, Juliette~T, Ziegenfuss, Jeanette~Y, Branda,
  Megan~E, Lim, Kaiser~G, Yawn, Barbara~P and Shah, Nilay~D}. (2012{\em b}).
\newblock The impact of asthma medication guidelines on asthma controller use
  and on asthma exacerbation rates comparing 1997--1998 and 2004--2005.
\newblock {\em Annals of Allergy, Asthma \& Immunology\/}~\textbf{108}(1),
  9--13.

\bibitem[Storlie \emph{and others}(2017)Storlie, Therneau, Carter, Chia,
  Bergquist and Romero-Brufau]{Storlie17a}
\textsc{Storlie, CB, Therneau, Terry, Carter, Rickey, Chia, Nicholas,
  Bergquist, John and Romero-Brufau, Santiago}. (2017{\em a}).
\newblock Prediction and inference with missing data in patient alert systems.
\newblock {\em Journal of the American Statistical Association (in review)\/}.
\newblock https://arxiv.org/pdf/1704.07904.pdf.

\bibitem[Storlie \emph{and others}(2013)Storlie, Michalak, Quinn, Dubois,
  Wender and Dubois]{Storlie12a}
\textsc{Storlie, Curtis~B, Michalak, Sarah~E, Quinn, Heather~M, Dubois,
  Andrew~J, Wender, Steven~A and Dubois, David~H}. (2013).
\newblock A {B}ayesian reliability analysis of neutron-induced errors in high
  performance computing hardware.
\newblock {\em Journal of the American Statistical
  Association\/}~\textbf{108}(502), 429--440.

\bibitem[Storlie \emph{and others}(2017)Storlie, Myers, Colligan, Weaver,
  Voigt, Croarkin, Leibson, Stoeckel, Katusic and Port]{Storlie17b}
\textsc{Storlie, Curtis~B., Myers, Scott, Colligan, Robert~C., Weaver, Amy~L.,
  Voigt, Robert, Croarkin, Paul~E., Leibson, Cynthia~L., Stoeckel, Ruth~E.,
  Katusic, Slavica~K. and Port, John~D.} (2017{\em b}).
\newblock Model-based clustering with mixed continuous and discrete variables
  via dirichlet process models.
\newblock {\em Biometrics (in review)\/}.
\newblock https://arxiv.org/pdf/1703.08741.pdf.

\bibitem[Tsiatis and Davidian(2001)Tsiatis and Davidian]{tsiatis2001}
\textsc{Tsiatis, Anastasios~A and Davidian, Marie}. (2001).
\newblock A semiparametric estimator for the proportional hazards model with
  longitudinal covariates measured with error.
\newblock {\em Biometrika\/}~\textbf{88}(2), 447--458.

\bibitem[Wulfsohn and Tsiatis(1997)Wulfsohn and Tsiatis]{wulfsohn1997}
\textsc{Wulfsohn, Michael~S and Tsiatis, Anastasios~A}. (1997).
\newblock A joint model for survival and longitudinal data measured with error.
\newblock {\em Biometrics\/}, 330--339.

\end{thebibliography}
\bibliographystyle{biorefs}
}

\label{lastpage}

\end{document}